\documentclass[12pt]{article}
\usepackage{graphicx}
\title{Content Distribution in Unicast Replica Meshes}
\author{ALEXEI NOVIKOV\\
Institute of Theoretical and Experimental Physics}
\begin{document}

\maketitle
\begin{abstract}
We propose centralized algorithm of data distribution in the unicast p2p
networks. Good example of such networks are meshes of WWW and FTP
mirrors. Simulation of data propagation for different network
topologies is performed and it is shown that proposed method performs up to 
200\% better then common approaches.
\end{abstract}
\section{Introduction}

One of the main methods of distribution of the free (open source)
software is via HTTP and/or FTP downloads. Such software is usually
replicated on different servers reducing load on individual servers and improving
client latency. Unfortunately some problems are associated with this method
of distribution, to name a few:
\begin{enumerate}\itemindent 10pt
\item Incomplete or outdated mirrors \cite{MT}
\item Absence of tools needed to locate nearest, best mirror server.
\item ``Slashdot-effect'' - After the new version of software is
announced the main server is overloaded for a long
period of time. This prevents proper update
of the mirror servers. 
\end{enumerate}

Intelligent replication of the software can solve first and the third
problem. Parallel download from the multiple mirror servers saturating the
available bandwidth can be used to resolve the second one.

Commercial content distribution networks (f.e. Akamai \cite{akamai})
are using IP multicasting for data transmission between replica nodes. 
Theirs networks are of the
order of hundred thousand hosts and both nodes and network
infrastructure are maintained by CDN.

Unfortunately IP multicast
suffers from a number of technical problems and the administrative
burden of maintenance. These problems prevent it from forming a global infrastructure. It can't
be widely used now as a transport layer for the global mesh of
replicas.  

On the other side there is a longstanding Internet convention that
large or popular sites are widely replicated, usually
voluntearly. This process is known as "mirroring". Software for
replication of WWW and FTP content exists for years - e.g. GNU wget
\cite{wget}, mirror \cite{mirror} and rsync \cite{rsync}. Recently
some formalization of this process has taken place with
organization of UK Mirror Service \cite{mirror_ac}, AARNet2 Mirror
Archive \cite{australia} etc. While the old volunteer based mirroring
services are operated on a best-effort basis, these new mirrors operate
under Service Level Agreements with guaranteed levels of
availability.

Mirror servers are updated using the standard client-server model
with client pulling data from the server. We propose a method of
content distribution in the server-server (or p2p) model that performs
much better then standard client-server approach.

In the next section we will review previous work done on the subject,
 in the third section will introduce our algorithm, in the
forth will describe the simulation setup, in section five we will
evaluate different heuristics and in the last section we will present
conclusions and formulate future development areas.

\section{Previous work}

Dispersity routing split the transfer of information over multiple network
paths to provide enhanced reliability and performance. One of the
first papers on the subject was the paper of
Maxemchuk \cite{Maxe}.
Usage of erasure codes in bulk data transfers
in multicasting environments was introduced in paper of Byers et all
\cite{Byers}. They used a notion of ``Digital Fountain'' - case when
servers are transmitting encoded parts of data independently, and
client can receive as much data as its bandwidth permits. 
One of the best known applications based on this idea is OpenCola
Swarmcast \cite{swarmcast}. 
It can be used as a content distribution media
but it was designed so that best performance is achieved at download of one huge file
at the presence of large number
of clients. Streaming media distribution based on the multiple description coding was
proposed in the work of Apostolopoulos and others
\cite{apostolopoulos}. They subdivided stream into 2
independent parts, each of them can be either played individually (with degraded
quality) or used to reconstruct initial file. These 2 streams can be
transmitted independently and from different servers.

The Yoid project \cite{yoid} was probably the best effort of creating
transport layer for a global CDN. Unfortunately it stoped at the stage of architectural
white paper.

A wast literature exists on replica placement in the Content
Distribution Networks (see f.e. \cite{kangasharju} and
\cite{radoslavov} and references therein). This problem is 
connected to the data
distribution in the p2p environments, and inspired this work.

To the best of our knowledge, this paper is the first one exploring
the problem of data distribution in the unicast networks in
many-to-many architecture.

\section{Content Distribution. Best-Client Selection Algorithm}

In this section we will introduce a simple model of data
distribution in the server-server network. Internet topology has been
studied in different papers \cite{List}-\cite{PowerLaw} since 1997. The main output
from these papers that is relevant to us, is that Internet
topology can not be modelled as a tree, but should be modelled as a
more complicated graph.  It was shown \cite{pas} that hop count
between Autonomous Systems (AS) is in good correlation with point-to-point
measured bandwidth. Papers \cite{mirsel} states that
performance of some of the mirror servers do not depend on this
property. That analysis can not be extended to the more
general case, due to the large number of valuable parameters not included in
the study.

Let assume that we do know the network topology $h_{ij}$ (hop count
between AS $i$ and AS $j$) and we do know
location of each of the $N$ nodes (replicas) in this network. Total
amount of data to be downloaded by each replica is $S$.
As soon as client downloads all data it becomes server and 
can start transmitting data to any other client on the network.

Let's assume that we have $n$ servers already.  Then we can write the
following proportionality for the bandwidth of the client $j$:

\begin{equation}
 b_i \sim \sum_{j \in n} \frac{1}{h_{ij}}.
\label{1}
\end{equation}

In  the capaciated version of this problem we have upper limits on
$b_i$ - bandwidth consumption during download  is limited by $u_i$
(client side) and bandwidth available from the server $j$ is limited by $d_j$
(server side). Then proportionality (\ref{1}) is transformed in to
following equation:

\begin{equation}
 b_i =min(\sum_{j \in n} min(f(\frac{1}{h_{ij}}),d_j), u_i),
\label{capac}
\end{equation}
where $f(\frac{1}{h_{ij}})$ is the hypothetical function giving point-to-point bandwidth
between nodes $i$ and $j$.

This equation provides us with 2 clues on what can be done in oder to maximize
bandwidth of the system at the shortest time:\\
a) client should be selected so that $b_i$ is maximal (in order to
increase number of servers as fast as possible).
\\b) download process should proceed from multiple servers to one client (in oder to reach
$u_i$) and to the multiple clients from one server (in oder to
reach $d_j$).

On one hand we can have not enough severs to saturate download limit of
all clients and
on another we can have not enough clients to consume all free bandwidth of
the servers. Then we can estimate
total bandwidth of the system to be

\begin{equation}
\sum_{i \in l}b_i =  min(\sum_{j \in n} d_j, \sum_{i \in l} u_i).
\label{sum}
\end{equation}

There are two distinct cases in the evolution of the system: \\a) when
$\sum_{j \in n} d_j \leq \sum_{i \in l} u_i$ (initial phase
) and \\b) $\sum_{j \in n} d_j \geq \sum_{i \in l} u_i$ (final phase).

For the simplicity let us assume that $d_j \equiv d$ and $u_i \equiv u$ for all hosts.
Then total bandwidth at time $t_0$ can be written as:

\begin{equation}
B(t_0)\approx d \;Int(\frac{\int^{t_0}_0B(t)dt}{S})+d \approx
d\;e^{\frac{d}{S}t_o}.
\label{begin}
\end{equation}

and

\begin{equation}
B(t_0) \approx u \;(N-Int(\frac{\int^{t_0}_0B(t)dt}{S})) \approx
\frac{ud}{u+d} N e^{-\frac{u}{S}(t_0-t^{\prime})},
\label{second}
\end{equation}

where $t^{\prime}=\frac{S}{d}\ln(\frac{uN}{d+u})$ - time when maximum
bandwidth is reached and amount of data servers can transmit per time 
equals to
amount of data clients can receive. Please note that integral only
roughly estimates amount of servers at $t_0$.

Another, and better, approximation of (\ref{second}) is 
\begin{equation}
B(t_0)=\frac{duN}{d+u},
\label{second_b}
\end{equation}
or bandwidth of the system at $t^{\prime}$.

Then $T$ - time  when the system will be filled, can be found from the
following equation:
 
\begin{equation}
 S\;N ={\int^T_0 B(t)dt},
\label{2}
\end{equation}
or
\begin{equation}
T\approx \frac{S}{d}\ln(\frac{uN}{d+u}) + \frac{S}{u}
\label{time}
\end{equation}

So we see that the time is proportional to the $S$ and that $u$ and $d$ should
be of the same order of magnitude. We should also try to
maximize $\sum_{j \in n} d_j$ or $b_i$ and $B$ at any given time.

Then the formal algorithm is the following: when there is any bandwidth
left on the server side, new client is selected so that $b_i$ is
maximal, or $\sum_{j \in n} h_{ij}$ is minimal. Clients can 
connect to the servers until they reaches $u_i$.

Let us provide a simple example clarifying why subdividing $S$ into 
smaller parts and transmitting them individually should give good
results.

We subdivide data into two equal parts and subdivide our replica
mesh into submeshes $A$ and $B$ so that mesh characteristics as the same.
Hop count distribution should be equal, and nodes nearest to the root server in each
group, should be located at the same distance. Then root server starts
sending parts $p_A$ and $p_B$ to the 2 nearest nodes in the meshes. Both
parts will be uploaded at $t_1\approx 2S/h$. After download is complete
these 2 sub-meshes can act as  independent networks, with
possible small degradation of connectivity. They will be
filled after $\approx T/2$ (see (\ref{time})). Then they will start filling each other:
if $d\leq u$ they will
fill each other at time $S/(2d)$, in other case that will happen at
$S/(2u)$. So the system will be  filled approximately 2 times faster compared to (\ref{time}).
Having in mind this example one can believe that time
of data propagation will be reduced after subdivision of data. We will
explore this approach in our simulation.

\section{Simulation}

For our simulation we choose generated networks of 100 nodes in 3 different
topologies. With the ``line'' topology we have 4 clusters of 25 nodes
that are randomly connected  within
cluster (with average distance of 4 hops).
Nodes from neighbour clusters are connected by 9 hops on
average, nodes that have one cluster in between-- 12 and 2 clusters -- 15 hops.

In ``triangle'' topology nodes within cluster are randomly connected
with average distance of 4, node is connected to any node in other
clusters with average distance of 12 hops.  

In ``random'' topology nodes are connected randomly with average
distance of 10.

All these topologies have average distance of 10 hops.

It was shown in \cite{PowerLaw} that
most of the
characteristics of the Internet structures (like hop distance between ASes
and router fanout) obey power-law. We will not use these empirical law in our
simulation, as replica placement can obey somewhat different laws  or
no laws at all.

Bandwidth between two nodes is inversely proportional to hop distance
with 25 \% of random noise  and varies from 1.5 MBps to 20 KBps.  

Amount of data to be distributed is 5 GB - around the size of modern Linux
distribution. 

Node selection is performed using algorithm proposed in previous
section. During simulation we include 5\% of random noise to bandwidth
between nodes. All simulated activity was subdivided in time slots of 1 minute,
during that interval nodes were only downloading and no node status
updates were made.

\subsection{Greedy-Global algorithm}

In oder to compare our results we choose algorithm used
in CDNs to locate position of replica/cache. This algorithm is called
Greedy-Global and with minimal modifications can be used in our
simulation. This modified method can be  formulated the following way:
We select replica one-by-one, at each step we choose one of the
nodes so, that if replica is placed there, the resulting network overhead
is minimized. 

Greedy Global heuristics can be modified to suit better p2p
distribution. Network overhead is characterized by the following cost
function:
\begin{equation}
C=\sum_{i}D_{i},
\label{GG}
\end{equation}
where sum runs over all client nodes and $D_i$ is minimal distance
from the client to the server.

Our algorithm is slightly different from
regular CDN GG algorithm, as request rate and object popularity are not
relevant in out case. 
This approach selects best location for $n$ replicas between $N$
nodes ( $n<<N$) under steady load from the clients. One would expect
that such algorithm can give good performance  in our model as well,
because when new node becomes server, our mesh have ``ideal'' replica
placement.

When the difference between cost  at step $k$ (k replicas) and at step
$k+1$ (k+1 replica)
becomes small we can assume that there are enough ``perfectly'' placed
replicas and we can start assigning clients based on our
Best Client approach in the submeshes belonging to each replica server. This
algorithm resembles current practice of assigning ``authoritative'' mirrors for each
geographical location and redirecting clients to them.

We will use both pure Greedy Global and Greedy Global/Best Client
algorithms in our simulation.

We can describe current ``best-practice'' method of data distribution in the
mesh of replicas as the
following procedure:
\begin{enumerate}\itemindent 10pt

\item Client tries to find the nearest mirror server with
required data and start download process.
\item If none is available, it
reaches master server.
\item If master server is full, it waits till master
server becomes free or some nearby server gets all data.
\end{enumerate}

We do not compare this method to ours, as according to our estimates,
it will perform worth then
Greedy Global.

\section{Performance evaluation}

Each test was run 100 times and data shown on figures is average.  
Different lines corresponds to 
data being subdivided into 2, 4, 8, 16 and 32 parts or not subdivided at
all.
\begin{figure}
\includegraphics[width=5.5cm,angle=270]{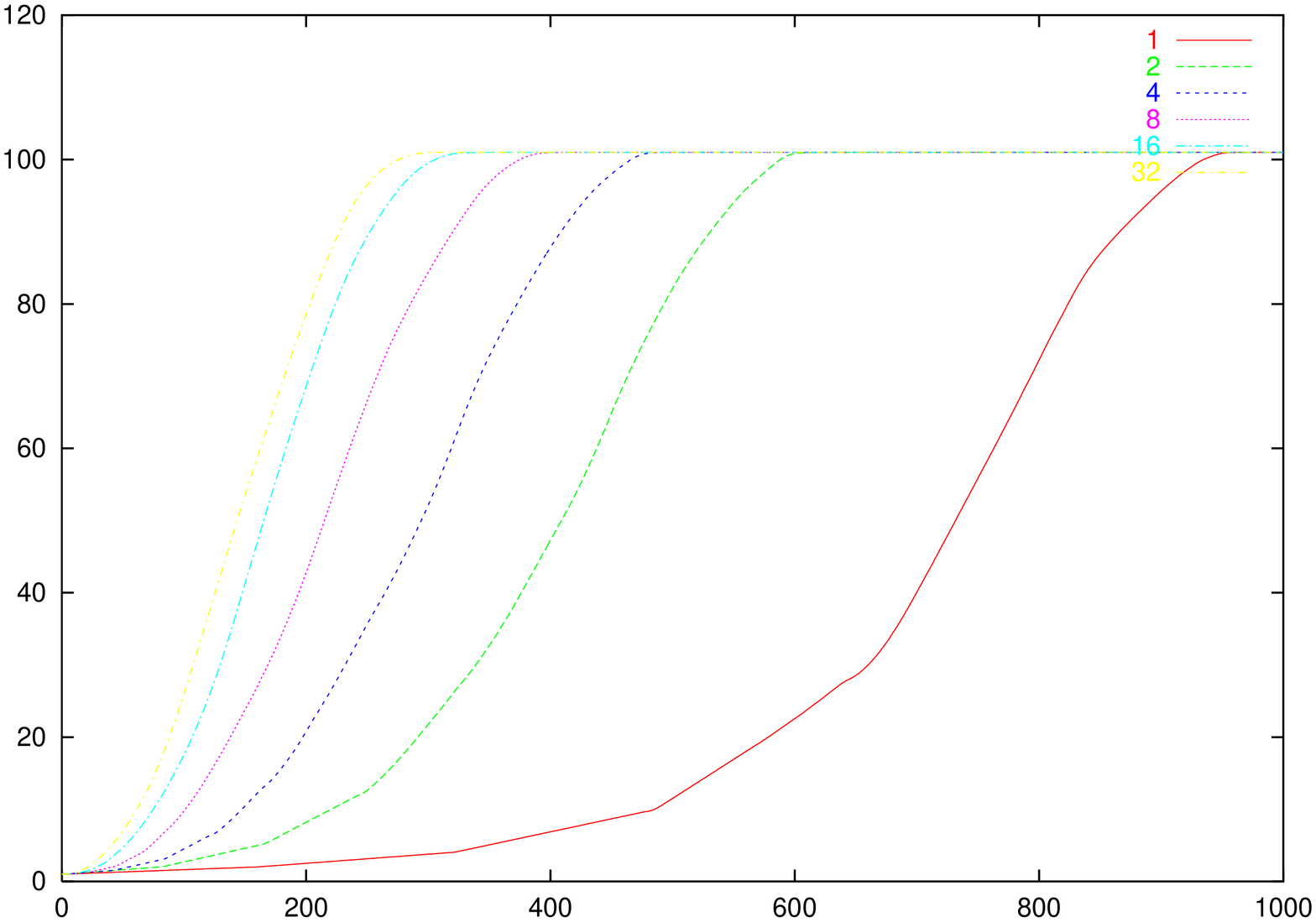}
\includegraphics[width=5.5cm,angle=270]{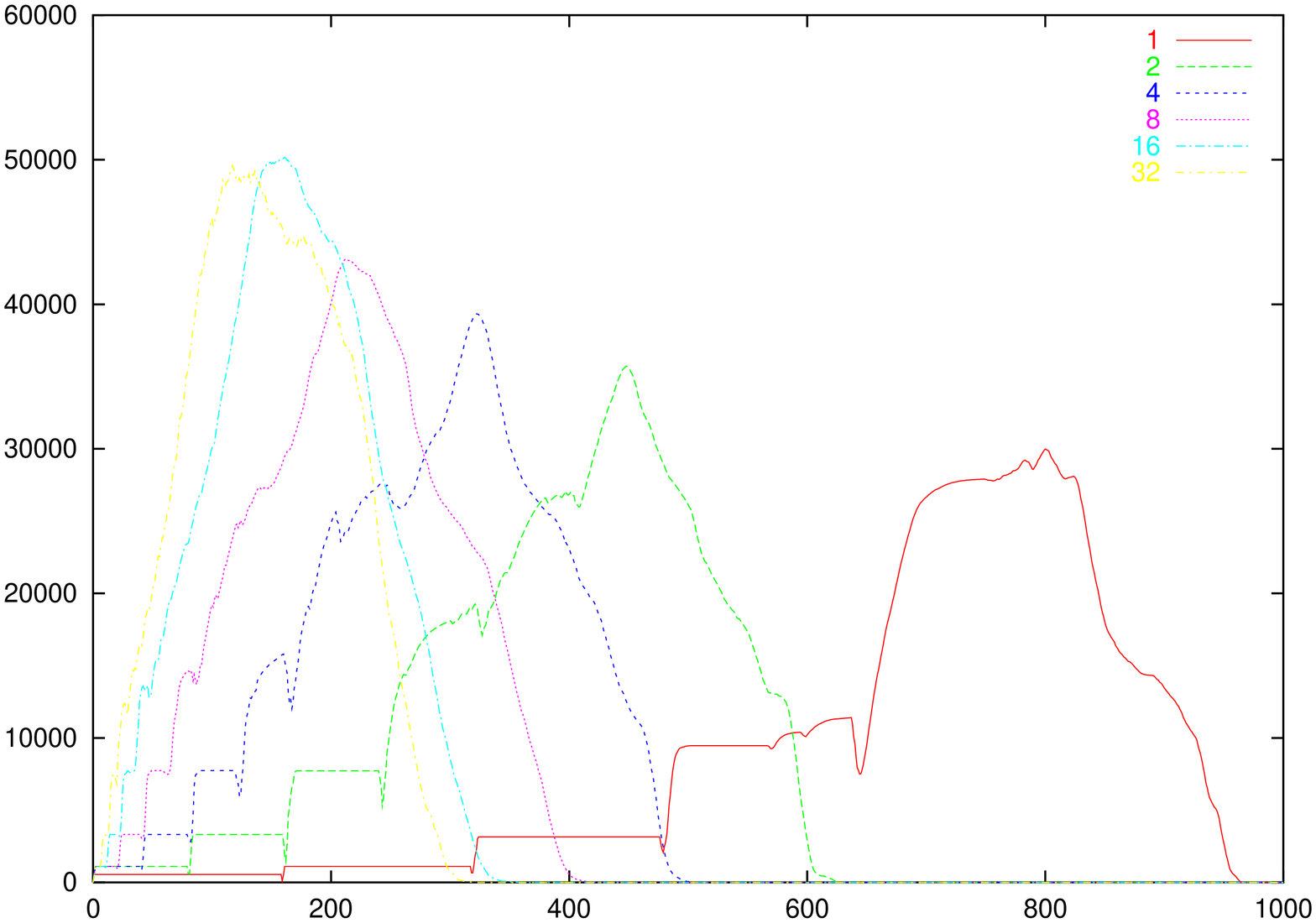}
\includegraphics[width=5.5cm,angle=270]{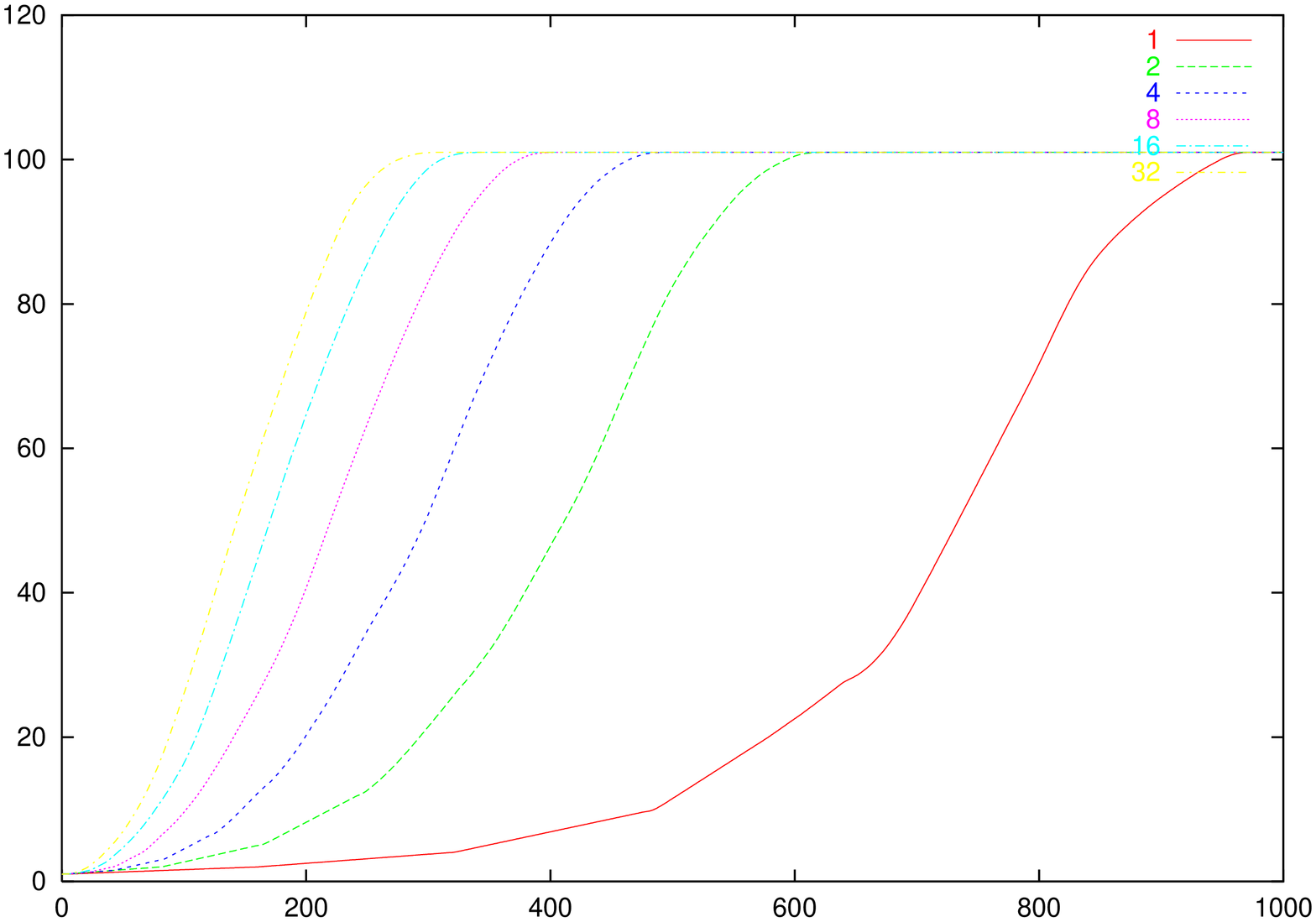}
\includegraphics[width=5.5cm,angle=270]{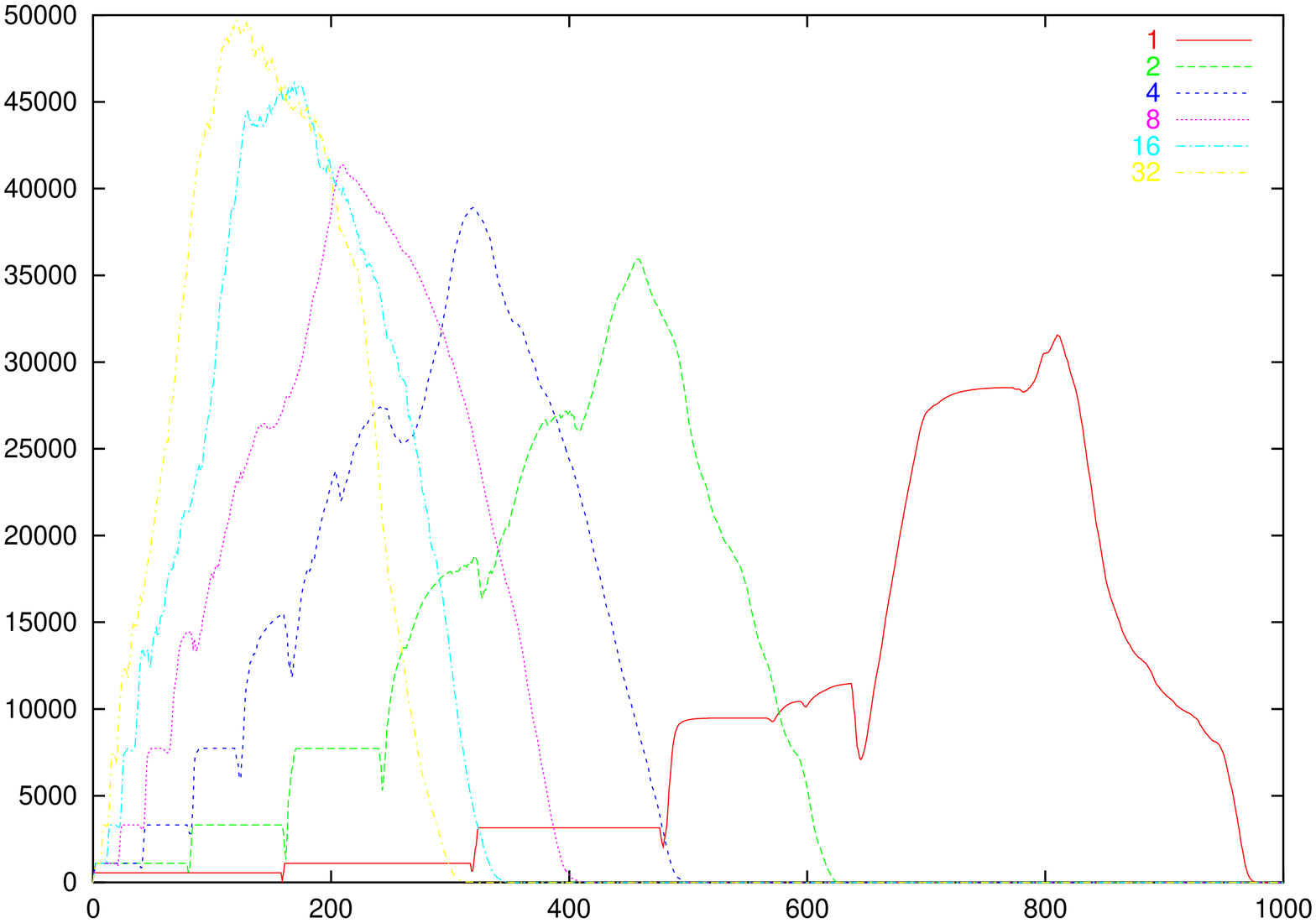}
\includegraphics[width=5.5cm,angle=270]{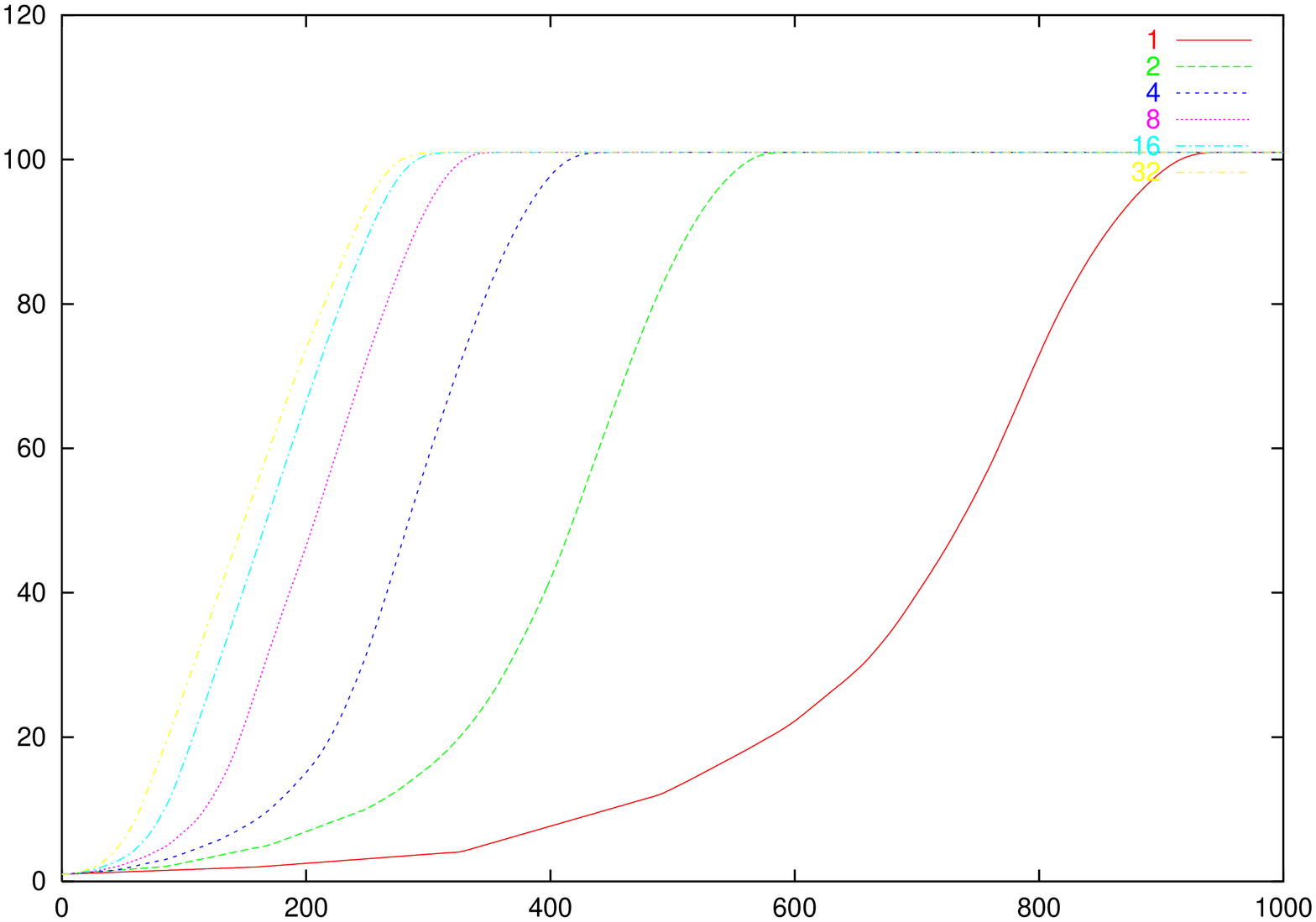}
\includegraphics[width=5.5cm,angle=270]{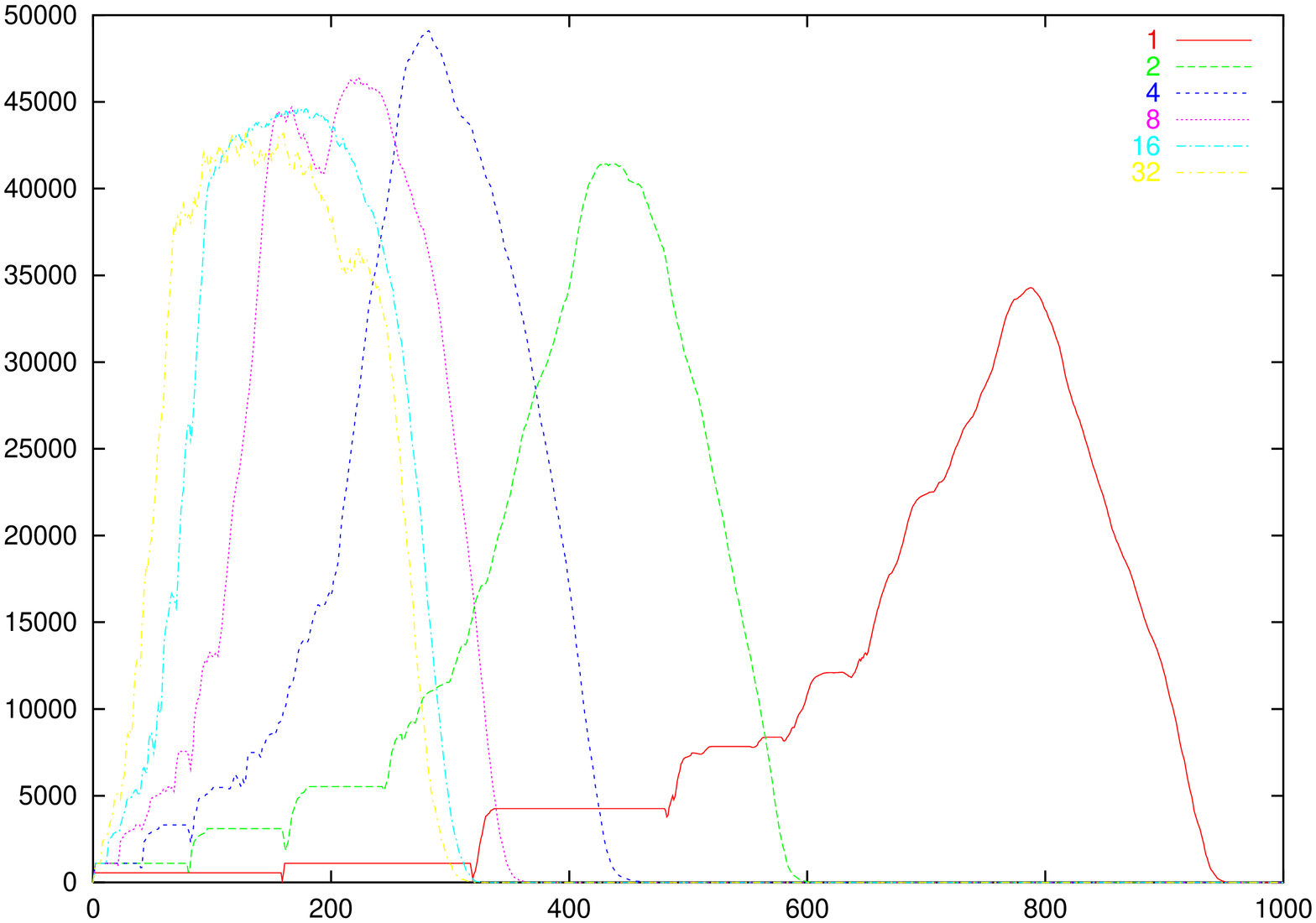}
\caption{Amount of nodes filled (left) and total system bandwidth $B$
  (right) vs time in ``line'', ``triangle'' and ``random'' topologies correspondingly}
\end{figure}

It can be easily seen from Figure 1 and Table 1, that subdividing data into
smaller chunks and transmitting them independently indeed improve
system performance.  Our results are almost independent on the
topology, this means that algorithm will
perform equally well in real-life. Time required to fill the system is
only slightly larger ($\approx 5\%$) for 16 then for 32 subparts for all
topologies. This can greatly reduce network overhead of the real system.

Results for the simulation with the Greedy Global algorithm are
presented on Figure 2.  For the sake of space we present figures only
for the ``line'' topology. As one can see the main disadvantage of this
heuristics is prolonged start of exponential growth. This is caused
by the first client selection. Usually the first selected client is
one with nearly the
worst connectivity to the root server, as opposed to the Best Client
method, where the first client is the nearest one. 

We provide more data on the time required to fill system in Table 1. Please
note that we include not only pure Greedy Global but also Greedy Global +
Best Client. We switch from Greedy to Best Client algorithm when the cost
of new placement  (\ref{GG}) is more then 70\%, 80\% and 90\% of the previous. 
\begin{figure}
\includegraphics[width=5.5cm,angle=270]{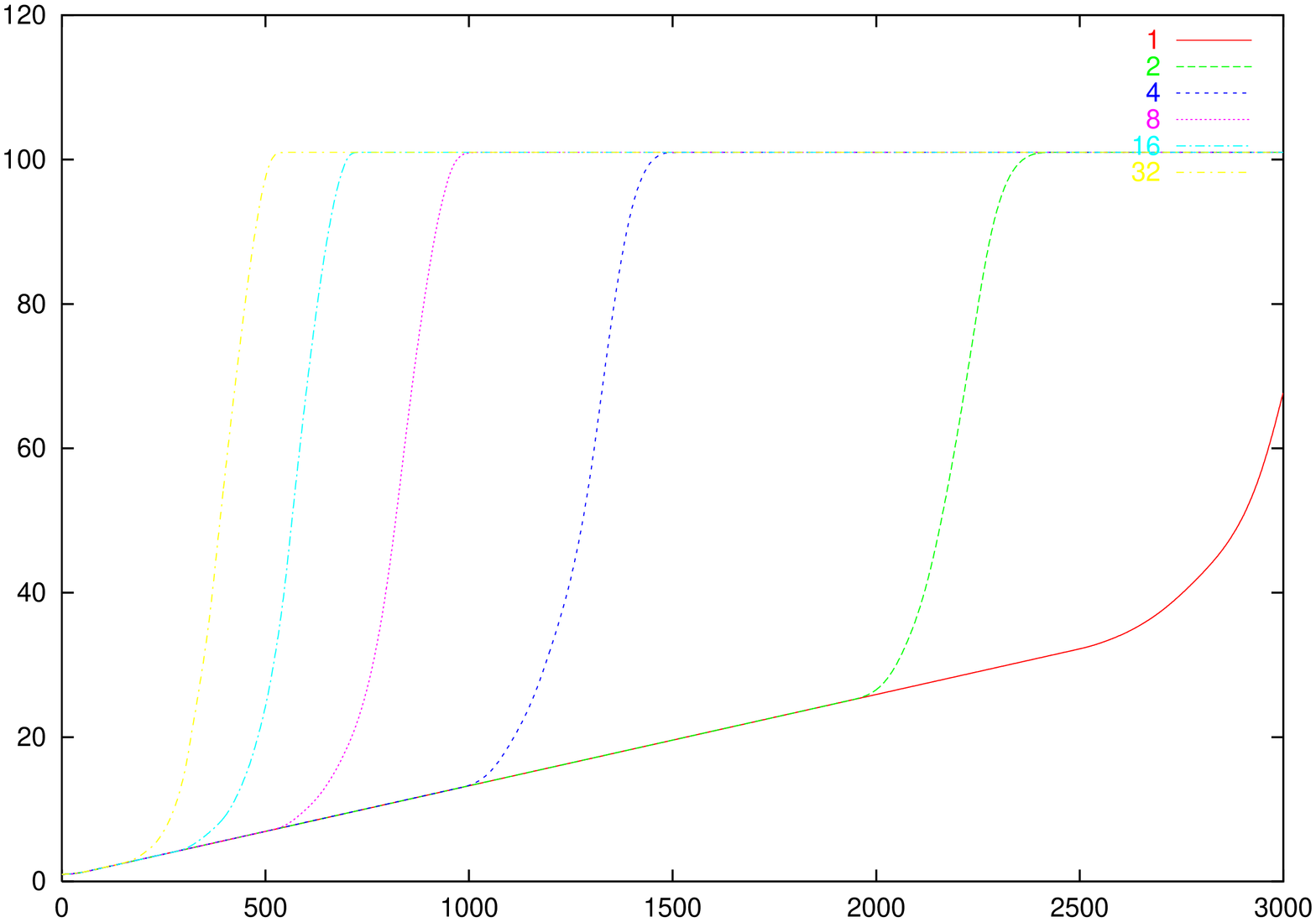}
\includegraphics[width=5.5cm,angle=270]{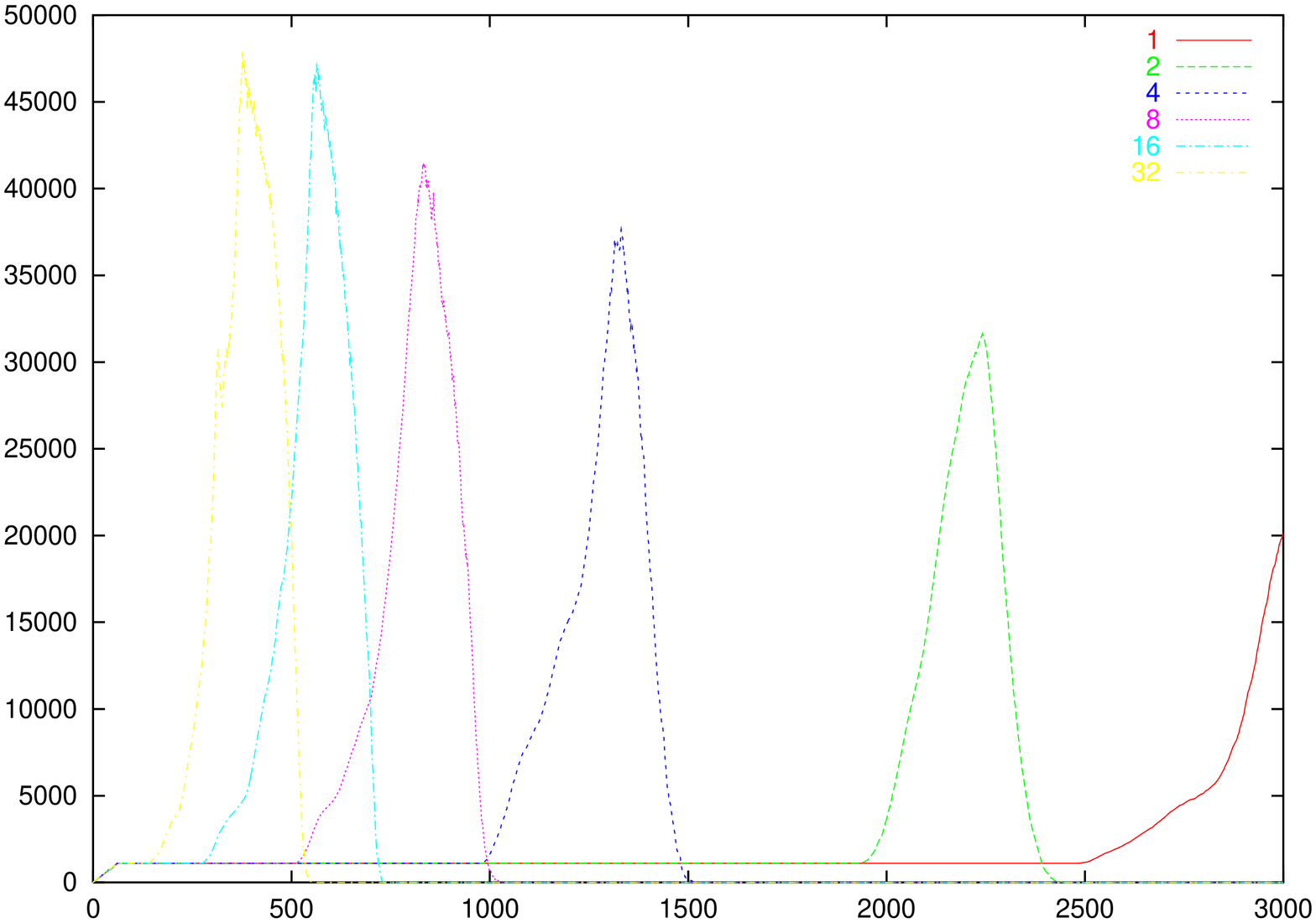}
\caption{Amount of nodes filled (left) and total system bandwidth $B$
  (right) vs time in ``line'' topology for the Greedy Global heuristics.}
\end{figure}
\newcommand{\mc}{\multicolumn}
\begin{table}
\centering
\(\begin{tabular}{|c|c|r|r|r|r|r|}
\hline
{Topology} & {\# of Parts}  & \mc{5}{|c|}{Time to fill (in min.)}\\
\cline{3-7}
{}&{} & {Best} & {Greedy} & {Greedy} & {Greedy
} & {Greedy}\\
{}&{} & {Client} & {Global} & {Global 0.9} & {Global 0.8} & {Global 0.7}\\
\hline
{} & {1} & {812(2)} & {$>$3000} &{$>$3000} & {$>$3000} & {$>$3000}\\
\cline{2-7}
{}&{2}&{622(4)} & {2420(40)} & {2724(276)} & {2726(274)}& {2516(43)}\\
\cline{2-7}
{}&{4}&{496(12)} & {1493(30)} &{1485(48)} & {1468(43)} & {1426(33)}\\
\cline{2-7}
{Line}&{8}&{407(17)} & {1000(24)}& {974(25)}&{973(32)} & {855(46)} \\
\cline{2-7}
{}&{16}&{347(12)} & {718(14)}& {666(30)}&{662(24)} & {576(13)}\\
\cline{2-7}
{}&{32}&{312(7)} & {540(13)} & {526(23)}&{493(26)}&{430(12)}\\
\hline
\hline
{} & {1} & {821(1)} & {$>$3000}& {$>$3000} & {$>$3000} & {823(6)}\\
\cline{2-7}
{Triangle}&{16}&{339(17)} & {657(12)}& {687(25)}& {530(25)}& {338(5)} \\
\cline{2-7}
{}&{32}&{323(5)} & {525(13)} & {516(31)} &{493(32)} & {291(7)}\\
\hline
\hline
{} & {1} & {825(20)} & {1202(96)} &{1684(300)} &{1338(38)} & {1122(60)}\\
\cline{2-7}
{Random}&{16}&{320(8)} & {493(25)}& {535(16)} &{530(25)} & {503(14)}\\
\cline{2-7}
{}&{32}&{311(8)} & {496(18)} & {505(27)}& {491(14)} & {474(8)}\\
\hline
\end{tabular}\)
\caption{Comparison of time required to fill the system for different heuristics, topologies, and
  number of subparts}
\label{tab:table}
\end{table}

One can easily see that Best Client outperforms all algorithms by
200\% for some topologies/parts and by 60\%  for another.

\subsection{Load balancing}

Our results can be quite sensitive to the load balancing policy.  We
looked at 4 different policies when selecting client with the same
cost function (\ref{1}) for the given chunk of data:
\begin{enumerate}\itemindent 10pt
\item Client that downloaded less of the given part.
\item Client that downloaded more of the given part.
\item Client that downloaded less totally.
\item Client that downloaded more totally.
\end{enumerate}

Results for different topologies and number of parts are presented in the
Table 2.  They show that time to fill is practically independent from the
load-balancing policy.  This means that current state of the system
can be known only to the level of node ( i.e. node being client or server for the
given part).
\begin{table}
\centering
\begin{tabular}{|c|c|r|r|r|r|}
\hline
{Topology} & {\# of Parts}  & \mc{4}{|c|}{Time to fill (in min.)}\\
\cline{3-6}
{}&{} & Policy 1. & Policy 2 & Policy 3 & Policy 4 \\
\hline
{} & {1} & {813(2)} & 812(2){} & {813(2)}& { 812(2)}\\
\cline{2-6}
{Line}&{16}&{341(5)} & {347(12)}&{348(9)} & {343(6)}\\
\cline{2-6}
{}&{32}&{310(8)} & {312(7)}&{313(9)} & {308(6)}\\
\hline
\hline
{} & {1} & {819(1)} & {821(1)}&{819(1)} & {821(1)}\\
\cline{2-6}
{Triangle}&{16}&{344(7)} & {339(17)}& {345(9)}& {336(8)}\\
\cline{2-6}
{}&{32}&{324(8)} & {323(5)}&{325(3)} & {328(9)}\\

\hline
\hline
{} & {1} & {824(11)} & {825(20)}&{824(11)} &{825(20)}\\
\cline{2-6}
{Random}&{16}&{320(6)} & {320(8)}& {322(5)}& {316(7)}\\
\cline{2-6}
{}&{32}&{308(4)} & {311(8)}& {310(7)}&{310(5)}\\
\hline
\end{tabular}
\caption{Comparison of time to fill for different topologies, 
  number of subparts and load balancing policies}
\end{table}
The
results presented on Figure 1 and Table 1 are given for the second
policy. 

\subsection{Free Loaders}

In our simulation all servers are serving as much data as
required. Having in mind good performance of our algorithm we can try
the capaciated version when all but root server are serving limited
amount of data (up to $1.2*S$). This case is common to the conventional
p2p networks when people are downloading but not serving data. This
people are called ``free loaders''. In order  to estimate impact of them we 
vary download limit from $5*S$
to $1.2*S$ and look at the time required to fill the system. Results are
presented in Table 3.
One can easily see that no reasonable performance degradation is
observed while mirror servers are providing at least twice the
original download.

\begin{table}
\centering
\begin{tabular}{|c|c|r|r|r|}
\hline
{Topology} & {\# of Parts}  & \mc{3}{|c|}{Time to fill (in min.)}\\
\cline{3-5}
{}&{} & 5 & 2 & 1.2\\
\hline
{} & {1} & {815(2)} & {989(8)} & {2264(81)}\\
\cline{2-5}
{Line}&{16}&{346(6)} & {345(7)}&{1242(120)}\\
\cline{2-5}
{}&{32}&{320(14)} & {315(10)}&{386(86)}\\
\hline
\hline
{} & {1} & {828(2)} & {1582(55)}&{2888(62)}\\
\cline{2-5}
{Triangle}&{16}&{343(9)} & {340(7)}& {891(89)}\\
\cline{2-5}
{}&{32}&{326(12)} & {332(10)}&{327(5)}\\

\hline
\hline
{} & {1} & {831(28)} & {1072(60)}&{1739(72)}\\
\cline{2-5}
{Random}&{16}&{318(6)} & {319(6)}& {324(7)}\\
\cline{2-5}
{}&{32}&{308(4)} & {308(6)}& {327(20)}\\
\hline
\end{tabular}
\caption{Comparison of time to fill for different topologies, 
  number of subparts and load balancing policies}
\end{table}

\section{Conclusions and future work}

We propose a method of
content distribution in the server-server (or p2p) model that performs
much better then standard client-server approach. 

Data to be downloaded from the server
is subdivided into given number of parts.
As soon as client downloads subpart completely it becomes server and 
can start transmitting it to any other client on the network. 
When there is any bandwidth left on the server side, 
new client is selected so that $b_i$ is maximal, 
or $\sum_{j \in n} h_{ij}$ is minimal. Clients can 
connect to the servers until they reaches $u_i$.

We perform simulation for different topologies and compare our
heuristic with modified Greedy Global and Greedy Global + Best Client
algorithms. We show that our method outperforms another algorithms by
200\% for some topologies/number of parts and by 60\%  for another.
Our results are almost independent on the
topology, so algorithm may perform equally well in real-life.

It was shown that subdividing data into
smaller chunks and transmitting them independently indeed improve
system performance.  
We also found that time required to fill the system is
only slightly larger ($\approx 5\%$) for 16 than for 32 subparts. 
Subdividing into smaller number of parts
can greatly reduce network and computational overhead of the real system.

We explore different load-balancing policies and find that time to
fill does not depend on them. This means that current
state of the system can be known only to the level of node (i.e. node
being client or server for the given part). 

We also tried to estimate impact of ``free-loaders''. In oder to do
that we varied download limit, and find that if servers are
transmitting at least twice the size of download, time required to fill
the system
remains almost the same. This means that our replica mesh may indeed
withstand ``slashdot effect''.

We are planing to develop decentralized algorithm that will use main
results of this paper. It will be interesting to specifically
check how many freeloaders can be in the mesh simultaneously and do
not degrade performance of decentralized algorithm.

One should check that our
system performs well in real Internet topology. 
For the completeness
we should also include simulation results for the current
``best-practice'' method of
data distribution in replica meshes.

\end{document}